\documentclass{PoS}

\makeatletter
\def\set@curr@file#1{%
  \begingroup
    \escapechar\m@ne
    \xdef\@curr@file{\expandafter\string\csname #1\endcsname}%
  \endgroup
}
\def\quote@name#1{"\quote@@name#1\@gobble""}
\def\quote@@name#1"{#1\quote@@name}
\def\unquote@name#1{\quote@@name#1\@gobble"}
\makeatother
\usepackage{graphics}

\newcommand{\beq}{\begin{eqnarray}}
\newcommand{\eeq}{\end{eqnarray}}
\newcommand{\nneeq}{\nonumber \end{eqnarray}}

\newcommand{\nn}{\nonumber \\}
\newcommand{\es}{& = &}

\newcommand{\rs}{\, = \,}
\newcommand{\ps}{& + &}

\newcommand{\np}{\nn \ps}

\newcommand{\cM}{ {\cal M} }
\newcommand{\cH}{ {\cal H} }

\newcommand{\2}{ \, {1 \over 2} \,}

\title{ Gauge boson mass as regulator of dynamics }

\ShortTitle{ Gauge boson mass as regulator }

\author{\speaker{Stanis{\l}aw D. G{\l}azek }\\
        Faculty of Physics, University of Warsaw\\
        E-mail: \email{stglazek@fuw.edu.pl}}

\abstract{Small-x divergences of Abelian gauge theory in the front form of Hamiltonian
   dynamics are regulated using a mass parameter for gauge bosons, introduced 
   through a mechanism analogous to the spontaneous breaking of global gauge 
   symmetry. A corresponding family of ultraviolet and infrared finite scale-dependent 
   renormalized Hamiltonians, is calculable order-by-order using the renormalization 
   group procedure for effective particles. The second-order terms described here
   suggest the magnitude of mass corrections that may be involved in resolving 
   the small-x parton and front-form vacuum and zero-mode problems, assuming 
   that the gauge boson mass that counts does not exceed the current upper bound 
   on the photon mass.}

\FullConference{Light Cone 2019 - QCD on the light cone: from hadrons to heavy ions - LC2019\\
		16-20 September 2019\\
		Ecole Polytechnique, Palaiseau, France}

\begin{document}

\section{The issue}

Front-form (FF)~\cite{DiracFF} Hamiltonians of quantum gauge 
theories (QGT) are singular. To become suitable for computation, 
they require regularization. To remove effects of regularization, 
they have to be renormalized. The question is how to do 
it~\cite{Wilsonetal}. Among other things, the issue is that in the 
FF dynamics of QGT the gauge bosons in gauge $A^+=0$ have 
polarization vectors 
\beq
\varepsilon_{p\sigma}^\mu \rs 
\left( \varepsilon^-_{p\sigma}
={2 p^\perp \varepsilon_\sigma^\perp \over p^+} , \ 
\varepsilon^+=0, \ \varepsilon_\sigma^\perp   \right)  \ ,
\eeq
where $\varepsilon^\perp_\sigma = (1+\sigma,1-\sigma)/2$ 
and $\sigma = \pm1$. The canonical minimal coupling of a current $j_\mu$ 
with the field $A^\mu$ has the form $j_\mu A^\mu$. This form 
leads to the Hamiltonian interaction terms $H_I$ that are proportional 
to $j^+\varepsilon_{p\sigma}^-$ and thus include the factor 
$p^\perp/p^+$. In the parton model~\cite{FeynmanPM}, 
the gauge boson carries a fraction $x$ of momentum $P$ of the 
system it belongs to and momentum $k$ that is transverse to $P$. 
In the FF dynamics, one has $ p^+ = xP^+$, $p^\perp =  x P^\perp 
+ k^\perp$. In perturbative description of the system, one encounters 
operator products such as $H_I (P^--H_f)^{-1} H_I$, where $H_f$ 
denotes the free Hamiltonian. For example, consider a fermion of 
momentum $P$ that emits and reabsorbs a gauge boson. The sum 
over the boson polarizations includes summing $|\varepsilon_{p\sigma}^-|^2$,
which yields $(k^\perp / x)^2$. The sum over intermediate fermion-boson 
systems involves integrating over the boson $x$ and $k^\perp$, for 
fixed $P$. The issue is that $(k^\perp/ x)^2$ makes the integral 
diverge for $x \to 0$ and $k^\perp \to \infty$. These divergences 
reinforce each other. 

How to handle the ultraviolet divergence due to large 
$k^\perp$ is understood~\cite{WilsonR}. A practical method to
handle small-$x$ divergences awaits invention. One can limit 
$p^+$ from below by a cutoff $\delta^+$. This excludes 
creation of field quanta from the bare vacuum state $|0\rangle$, 
which can thus be used to build a whole Fock space of states in 
which the canonical Hamiltonian acts. However, $\delta^+$ 
breaks boost invariance and one may prefer the cutoff $x > 
\delta$, where $\delta$ is a pure number that determines 
the minimal value of $x$ irrespective of the value of $P^+$. 
We proceed in a different way, focusing on Abelian theory 
as the simplest one that exhibits $(k^\perp / x)^2$ behavior.

We apply the renormalization group procedure for effective
particles (RGPEP)~\cite{RGPEP}. Small-$x$ regularization 
results from introducing mass for the gauge bosons. The mass 
is introduced via a mechanism analogous to the spontaneous 
breaking of global gauge symmetry~\cite{Higgs1,EnglertBrout}
and we work in the limit that yields Soper's FF version of massive 
QED~\cite{Soper,AbelianAPPB}. Massive gauge theories have 
a history of studies using light-front quantization methods with 
various regularizations, see~\cite{Hiller} and references therein. 
We address the issue of evaluating finite scale-dependent and 
boost invariant effective Hamiltonian operators. Specifically, we 
describe the mass corrections and discuss the orders of magnitude 
of terms that one needs to deal with. The heuristic value of such 
consideration is that the RGPEP studies can order-by-order (in 
expansion in powers of a coupling constant) teach us about the 
FF dynamics of gauge field quanta. This discussion concerns 
only terms of second order in Abelian Soper's theory.  

\section{ RGPEP }

The canonical Hamiltonian of a QGT, denoted by $H_{\rm can}$, 
is considered an initial condition for solving the differential 
equation ($\cH_f$ is the free part of $H_{\rm can}$ and $\tilde \cH_t$
is simply related to $\cH_t$, {\it cf.}~\cite{RGPEP})
\beq
\label{Hprime}
\cH'_t \es [  [ \cH_f, \tilde \cH_t ] , \cH_t ] \ ,
\eeq
for $\cH_t$ so that $\cH_{t=0} = H_{\rm can}$. The equation 
originates from the similarity renormalization group 
procedure~\cite{GlazekWilson} and results from adapting
the double-commutator flow equation for Hamiltonian 
matrices~\cite{Wegner} to the purpose of calculating
effective FF Hamiltonian operators $H_t$. 
Solving Eq.~(\ref{Hprime}) for the Hamiltonians $\cH_t$ 
is an intermediate step. They are polynomials in creation 
and annihilation operators of canonical theory. The 
polynomial coefficients, say $c_t$, that are found by 
solving Eq.~(\ref{Hprime}), are used to build the 
Hamiltonians $H_t$ for effective particles. Namely, 
$H_t$ is obtained by replacing canonical creation and 
annihilation operators in $\cH_t$ by the ones for 
effective particles that are labeled by $t$. Quantum 
numbers are the same. The coefficients $c_t$ depend 
only on these quantum numbers and they are the same 
in $\cH_t$ and $H_t$. The unitary relationship between 
the canonical and effective particle operators is the source 
of Eq.~(\ref{Hprime})~\cite{RGPEP}. One works with 
operators instead of matrix elements and the generator 
of the unitary transformation is designed to preserve all 
seven kinematic Poincar\'e symmetries of the FF of 
Hamiltonian dynamics. Divergences of the theory appear 
in $H_t$ in explicit form. One deals with them by adding 
counter terms to the canonical Hamiltonian $H_{\rm can}$. 
We describe the magnitude of mass counter terms and 
effective mass corrections one obtains in Soper's theory~\cite{Soper}.

\section{ Second-order mass corrections } 

Applying the RGPEP to Soper's theory, one starts from its
FF Hamiltonian density~\cite{Soper,AbelianAPPB}, 
\beq
\label{cH16}
\cH \es
\bar \psi_f \gamma^+ { (i \partial^\perp)^2 + m^2 \over 2
i\partial^+} \psi_f
+
\2 A_f^i \left[ (i\partial^\perp)^2 + \kappa^2 \right] A_f^i 
+
\2 \ B \left[ (i \partial^\perp )^2 +  \kappa^2 \right] B 
\np
g \bar \psi_f \not \hspace{-4pt} A_f \psi_f 
-
g \bar \psi_f \gamma^+ \psi_f  {\kappa \over i\partial^+ } iB
+
\2 g^2 \bar \psi_f \not \hspace{-4pt} A_f { \gamma^+ \over i
\partial^+} \not \hspace{-4pt} A_f \psi_f
+
\2 \left[ {1 \over \partial^+ } g \bar \psi_f \gamma^+ \psi_f
\right]^2 \ ,
\eeq
applies the standard light-front quantization procedure and 
solves Eq.~(\ref{Hprime}). The density includes the fermion 
field $\psi$, transverse boson field $A$ with two polarizations 
in gauge $A^+=0$ and an additional gauge boson field $B$,
associated with the third polarization state of massive vector 
bosons. 

Regularization originates from the RGPEP form factors that result
from solving Eq.~(\ref{Hprime}) for terms order $g$. They have 
the form $\cH_{t 1 \, c.a} = \exp[-t (\cM_c^2 - \cM_a^2)] \  \cH_{0 1 \, c.a}$,
where $c$ and $a$ refer to operators that create and annihilate
quanta and $\cM$ denotes their total invariant mass, correspondingly. 
One multiplies the bare interaction terms by the form factor 
with $t$ replaced by $t_r$. Singularities due to $(k^\perp/x)^2$
are regulated because the mass $\kappa$ enters $\cM^2$ through
$(k^{\perp 2} + \kappa^2)/x$. The regularization is lifted when $t_r \to 0$.

In the series expansion $H_t = H_{t f} + g H_{t1} + g^2 H_{t2} 
+ O(g^3) $, the bare coupling constant $g$ differs from the effective 
coupling constant $g_t$ first in third order. We do not need to 
distinguish them here, because our discussion only concerns the 
mass squared terms in $H_t$ that include the free terms from 
$H_{t f}$ and second-order mass squared corrections from 
$g^2 H_{t2}$. Namely,
\beq
\label{Hpsi2}
H_{t \, \psi} 
\es 
\sum_{\sigma = 1}^2 \int [p] \ {p^{\perp \, 2} + m^2 + g^2 \delta m^2(t) \over p^+} \
  \left[b^\dagger_{t \, p\sigma }b_{t \, p\sigma } + 
  d^\dagger_{t \, p\sigma }d_{t \, p\sigma } \right] \ , \\
\label{HA2}
H_{t \, A} 
\es 
\sum_{\sigma =1}^2 \int [p] \ {p^{\perp \, 2} + \kappa^2 + g^2 \delta \kappa_A^2(t) \over p^+} \
  a^\dagger_{t \, p\sigma}a_{t \, p\sigma} \ , \\
\label{HB2}
H_{t \, B} 
\es 
\int [p] \ {p^{\perp \, 2} + \kappa^2 + g^2 \delta \kappa_B^2(t) \over p^+} \
  c^\dagger_{t \, p} c_{t \, p} \ .
\eeq
The mass-squared counter terms in the canonical Hamiltonian
are adjusted so that the eigenvalue problems for a single
physical particle have eigenvalues $m^2$ for fermions and 
$\kappa^2$ for bosons of types $A$ and $B$. After integration 
over $k^\perp$,
\beq
\label{deltamplot}
g^2 \delta m^2(t) 
\es
{\alpha_g \over 4 \sqrt{2\pi} } 
\ 
{I_1(t) \over \sqrt{t+t_r}} 
-
{\alpha_g \over 4 \pi}
\left( 2 m^2 + \kappa^2 \right) 
\ 
I_2(t) 
\ , \\
\label{kappaAplot}
g^2 \delta \kappa_A^2(t) 
\es
{\alpha_g \over 4 \sqrt{2\pi} } 
\ 
{I_3(t) \over \sqrt{t+t_r}} 
+
{\alpha_g \over 4\pi } 
\left( 2 m^2 + \kappa^2 \right) 
\ 
I_4(t)
\ , \quad
g^2 \delta \kappa_B^2(t) 
\rs
 {\alpha_g \over 4 \pi } \ \kappa^2 
\
I_5(t)
\ ,
\eeq
where $\alpha_g=g^2/(4\pi)$ and the scale-dependent integrals are
\beq
\label{IFE}
I_1(t) 
\es \int_0^1 dx \
{1 + (1-x)^2 \over x }   \ {\rm erfc}\left[ \sqrt{2(t+t_r)} \, \delta \cM_{fb}^2 \right] \ , \\
\label{IFG}
I_2(t) 
\es \int_0^1 dx \
\Gamma \left[0, 2(t+t_r) \, \delta\cM_{fb}^4 \right] \ , \\
\label{IAE}
I_3(t) 
\es \int_0^1 dx \
\left[  x^2 + (1-x)^2 \right]  \ {\rm erfc} \left[ \sqrt{2(t+t_r)}\, \delta \cM_{f \bar f}^2 \right] \ , \\
\label{IAG}
I_4(t)
\es \int_0^1 dx \
\left[ 1 - { \kappa^2 x(1-x)  \over m^2 + \kappa^2/2 }\right]  
\ \Gamma \left[ 0,2(t+t_r) \, \delta \cM_{f \bar f}^4 \right] \ , \\
\label{IBG}
I_5(t)
\es \int_0^1 dx \ 4 x(1-x) \
\Gamma \left[ 0,2(t+t_r) \, \delta \cM_{f \bar f}^4 \right] \ .
\eeq
erfc and $\Gamma$ are the complementary error and 
incomplete gamma functions.  Their arguments include
$\delta \cM_{fb}^2 = \kappa^2/x +  m^2/(1-x) - m^2$
and $\delta \cM^2_{f \bar f} = m^2/x + m^2/(1-x) - \kappa^2$.
In the limit $t \to 0$, Eqs.~(\ref{deltamplot}) and (\ref{kappaAplot}) 
provide the values of the mass-squared counter terms introduced 
in the initial, canonical Hamiltonian that is regulated using $t_r$. 

The effective fermion mass correction behaves like 
$- \ln (\kappa^2 \sqrt{t+t_r})/ \sqrt{t + t_r}$ for small $t$. 
The boson $A$ and $B$ corrections are less singular but
they significantly differ from each other, though the physical masses, 
or eigenvalues of the Hamiltonians $H_t$ for all finite values 
of $t$, are equal in second order calculation $m^2$ for 
the fermions and $\kappa^2$ for the both types of bosons, 
$A$ and $B$. 

\section{ Orders of magnitude } 

We set $\alpha_g = 1/137$ and provide examples of the mass corrections we 
\begin{table}[ht!]
  \begin{center}
      \label{tab:table1}
    \begin{tabular}{|c|c|c|c|c|c|c|}
\hline
$\kappa = m \quad s \, m:                              $ & $ 1                     $ & $ 0.5                  
                                         $ & $ 0.25                 $ & $ 0.1                   
                                         $ & $ 0.01                 $ & $ 0.001               $ \\      
\hline
$g^2\delta m^2/m^2         $ & $ 3.19 \ 10^{-13} $ & $ 3.03 \ 10^{-4}                   
                                         $ & $ 1.88 \ 10^{-2}   $ & $ 3.67 \ 10^{-1}   
                                         $ & $ 1.04 \ 10^{2}    $ & $ 1.71 \ 10^{4}  $ \\
\hline
$g^2\delta \kappa_A^2/m^2  $ & $ 2.73 \ 10^{-13} $ & $ 1.69 \ 10^{-4} 
                                             $ & $ 5.17 \ 10^{-3}  $ & $ 4.67 \ 10^{-2}   
                                             $ & $ 4.85                $ & $  4.85 \ 10^{2}   $ \\
\hline   
$g^2 \delta \kappa_B^2/m^2 $ & $ 5.64 \ 10^{-14} $ & $ 3.38 \ 10^{-5}                 
                                             $ & $ 6.21 \ 10^{-4}   $ & $ 1.96  \ 10^{-3}  
                                             $ & $ 5.52 \ 10^{-3}   $ & $ 9.09 \ 10^{-3}    $ \\
\hline         
    \end{tabular}
  \end{center}
\end{table}
obtain for $\kappa = m$. This case qualitatively illustrates 
the situation one encounters in commonly considered systems 
made of two constituents of comparable masses. The parameter 
$s = t^{1/4}$ can be intuitively understood as the size of 
effective quanta. The mass corrections are indeed small for 
the size $s$ comparable or larger than the fermion Compton 
wavelength. However, they quickly grow when $s$ becomes 
smaller than the wavelength. 

The mass corrections greatly increase when one considers the 
boson mass $\kappa$ comparable to the experimental upper 
limit on the photon mass, $m_\gamma < 10^{-18}$ eV~\cite{PDG}. 
The mass corrections we obtain for the corresponding $\kappa=
10^{-25} \, m$, assuming $m$ is on the order of the electron 
mass, are given in the table below. 
\begin{table}[ht!]
  \begin{center}
        \label{tab:table2}
    \begin{tabular}{|c|c|c|c|c|}
\hline
$\kappa=10^{-25} \, m \ , \quad s \, m :                                 $ & $ 1                        $ & $ 10^{-1}                 
                                            $ & $ 10^{-3}              $ & $ 10^{-6}             $ \\      
\hline
$g^2\delta m^2/m^2            $ & $ 1.62 \ 10^{-1}    $ & $ 1.71  \  10^{+1}                  
                                            $ & $ 1.85 \ 10^{+5}   $ & $ 2.05 \ 10^{+11}  $ \\
\hline
$g^2\delta \kappa_A^2/m^2 $ & $ 1.39 \ 10^{-4}    $ & $ 4.93  \ 10^{-2} 
                                            $ & $ 4.85 \ 10^{+2}   $ & $ 4.85  \ 10^{+8}  $ \\
\hline   
$g^2 \delta \kappa_B^2/m^2 $ & $ 3.17 \ 10^{-70}  $ & $ 1.79  \ 10^{-53}                 
                                             $ & $ 8.92 \ 10^{-53}  $ & $ 1.96  \ 10^{-52} $ \\
\hline         
    \end{tabular}
  \end{center}
\end{table}
Such large corrections are less harmful to the theory than one might 
expect because the mass corrections are exactly canceled by the 
self-interactions of effective particles and the eigenvalues continue 
to not depend on $t$. However, the magnitude of terms that cancel 
out can be very large unless one keeps the size $s$ of effective particles 
in the right range. This way the mass correction hierarchy problem is 
resolved by the finite size of quanta.

\section{ Conclusion }

The RGPEP computation of effective mass corrections in Soper's theory 
suggests a path to take for calculating other Hamiltonian interaction terms
in particle theory. Local gauge symmetry with minimal coupling approximates 
effective interactions of fermions for invariant mass changes much smaller 
than the inverse of their Compton wavelength. The hierarchy problem 
for mass squared corrections is similarly resolved. The author hopes to 
discuss non-Abelian gauge theories with infinitesimally spontaneously 
broken global gauge symmetry for regularization purposes in a near 
future elsewhere.



\begin{thebibliography}{99}

\bibitem{DiracFF}
P. A. M. Dirac, 
Rev. Mod. Phys. {\bf 21}, 392 (1949).

\bibitem{Wilsonetal}	
K. G. Wilson et al., 
Phys. Rev. D {\bf 49}, 6720 (1994).

\bibitem{FeynmanPM}
R. P. Feynman,
Phys. Rev. Lett. {\bf 23}, 1415 (1969).

\bibitem{WilsonR} 
K. G. Wilson,
Phys. Rev. D {\bf 2}, 1438 (1970).

\bibitem{RGPEP}
S. D. G{\l}azek,
Acta Phys. Polon. B {\bf 43}, 1843 (2012).

\bibitem{Higgs1}
P. W. Higgs,
Phys. Lett. {\bf 12}, 132 (1964).

\bibitem{EnglertBrout}
F. Englert, R. Brout,
Phys. Rev. Lett. {\bf 13}, 321 (1964).

\bibitem{Soper}
D. E. Soper, Phys. Rev. D {\bf 4}, 1620 (1971).

\bibitem{AbelianAPPB}
S. D. G{\l}azek, 
Acta Phys. Polon. B {\bf 50}, 5 (2019).

\bibitem{Hiller}
J. R. Hiller, 
Prog. Part. Nucl. Phys. {\bf 90}, 75 (2016).

\bibitem{GlazekWilson}
S. D. Głazek, K. G. Wilson, 
Phys. Rev. D {\bf 48}, 5863 (1993).

\bibitem{Wegner}
F. Wegner, Ann. Phys. (Leipzig) {\bf 3}, 77 (1994).

\bibitem{PDG}
M. Tanabashi et al. (Particle Data Group), 
Phys. Rev. D {\bf 98}, 030001 (2018).

\end{thebibliography}
\end{document}